\documentclass[fleqn,10pt,dvipsnames]{wlscirep}
\usepackage[utf8]{inputenc}
\usepackage[T1]{fontenc}

\usepackage{amsmath}
\usepackage{physics}
\usepackage{svg}
\usepackage{float}
\usepackage{soul}
\usepackage[flushleft]{threeparttable}

\title{A Quantum Hopfield Associative Memory Implemented on an Actual Quantum Processor}

\author[1,*]{Nathan Eli Miller}
\author[1]{Saibal Mukhopadhyay}
\affil[1]{Georgia Institute of Technology, School of Electrical and Computer Engineering, Atlanta, GA, 30332, USA}

\affil[*]{nathan.miller@gatech.edu}

\begin{abstract}
In this work, we present a Quantum Hopfield Associative Memory (QHAM) and demonstrate its capabilities in simulation and hardware using IBM Quantum Experience. The QHAM is based on a quantum neuron design which can be utilized for many different machine learning applications and can be implemented on real quantum hardware without requiring mid-circuit measurement or reset operations. We analyze the accuracy of the neuron and the full QHAM considering hardware errors via simulation with hardware noise models as well as with implementation on the 15-qubit \textit{ibmq\_16\_melbourne} device. The quantum neuron and the QHAM are shown to be resilient to noise and require low qubit overhead and gate complexity. We benchmark the QHAM by testing its effective memory capacity and demonstrate its capabilities in the NISQ-era of quantum hardware. This demonstration of the first functional QHAM to be implemented in NISQ-era quantum hardware is a significant step in machine learning at the leading edge of quantum computing.
\end{abstract}
\begin{document}

\flushbottom
\maketitle

\thispagestyle{empty}

\section*{Introduction}

Since the advent of quantum computing, one of the primary applications which has piqued the interest of both academia and industry is quantum machine learning.\cite{IntroQML,NearTerm,QML,Google} With the characteristics of quantum computing driven by superposition and entanglement of quantum states, quantum computing can create and model patterns in data which cannot be easily created in classical computing. Additionally, quantum parallelism allows for quantum gates to evaluate outputs of functions with multiple inputs simultaneously.\cite{SimpleQuantumComputer} Quantum machine learning is therefore expected to produce significant speedup in machine learning algorithms.\cite{QML} Already, quantum machine learning models have been developed for generic quantum neural networks,\cite{GG_QN,GenericQNNs,PeriodicQNN} quantum dynamic neural networks,\cite{QDNN} continuous-variable quantum neural networks,\cite{ContQNN} quantum convolutional neural networks,\cite{ConvQNN,Quan} and quantum Hopfield neural networks.\cite{QHN} In large part, quantum machine learning research has recently been centered around the use of variational quantum algorithms as machine learning models \cite{VariationalQAlgos,QCModelsForANNs,NISQAlgosBharti,AmiraPowerQNNs} due to their high expressibility and rapid trainability compared to classical networks.\cite{AmiraPowerQNNs} Extensive research has also been done at the intersection of classical and quantum machine learning to create hybrid machine learning techniques,\cite{HybridNN,HybridCNN,QAutoEn} as well as in techniques for training quantum neural networks \cite{TrainingDQNN} and in machine learning using quantum-enhanced feature spaces.\cite{QEFS}

The characteristics of quantum computing are of particular interest for Hopfield network \cite{Hopfield} applications.\cite{QHN,GG_QN,exponential_qham,qnn_quest} Hopfield networks become intractable to compute classically due to their fully connected nature, though a recent paper claims ``Hopfield networks is all you need'' \cite{AllYouNeed} for many complex machine learning tasks. With quantum computing, significant improvements in time and memory overhead for Hopfield networks have been claimed through algorithmic advantages, training from a superposition of input states and parallel computation,\cite{QHN,GG_QN} resulting in theorized polynomial or exponential capacity improvements for some quantum associative memory designs.\cite{exponential_qham,qnn_quest,QuAM}

Several studies theorize quantum associative memories with improved capacity,\cite{QuAM,exponential_qham} though neither have been implemented in hardware. The original design in 1998 \cite{QuAM} utilizes both spin-$\frac{1}{2}$ and spin-1 qubits to demonstrate a theoretical associative memory, and a more recent design \cite{exponential_qham} involves long-range interactions of fully-connected qubits to generate a unitary evolution over time. Another approach implements a quantum neuron to perform pattern recognition,\cite{NatureNeuron} but this design does not follow Hopfield dynamics and has not been extended to an associative memory system.

In the NISQ era of quantum computing,\cite{NISQ} quantum computers are limited by their low number of qubits and their high vulnerability to noise and quantum errors. Therefore, implementation of quantum machine learning architectures in quantum hardware is very difficult. Of the above designs, only one has been implemented in hardware.\cite{NatureNeuron}

In this work, we demonstrate a Quantum Hopfield Associative Memory (QHAM) implementable in IBMQ,\cite{IBMQ} IBM's open source development platform allowing for the implementation of quantum algorithms in quantum hardware based on superconducting qubits. We consider the effects of quantum errors in simulation via the IBMQ QASM simulator,\cite{IBMQ_qasm} as well as in IBMQ hardware devices of up to 15 qubits. We differentiate from prior works by following Hopfield dynamics with the use of a quantum neuron and by implementing the QHAM in a format compatible with quantum gate-based circuits utilized in IBMQ. Our design can be tested on an actual quantum processor in IBMQ considering noise sources such as readout noise,\cite{noise} quantum gate errors, and qubit decoherence.\cite{Decoherence} Both our quantum neuron and our QHAM show resilience to noise and do not require mid-circuit measurement or reset operations. Thus, our QHAM implementation is the first QHAM to be successfully demonstrated in quantum hardware. 

\section*{Results}

\subsection*{Background on Hopfield Associative Memories}

The Hopfield network, first developed by J. J. Hopfield in 1982,\cite{Hopfield} is a type of classical neural network which has demonstrated widespread capabilities in machine learning, most notably in creating associative (content-addressable) memories. In an associative memory, one or more memory states are learned by the network and stored in the network's weight matrix $w$ (also called the neural interaction matrix). These states can be recovered by introducing an input state to the network which is close in Hamming distance to one of the stored states, then evolving the state of the network by updating individual neurons until convergence is reached.

Consider a Hopfield network of $n$ fully connected neurons. Let each memory state stored in the network be represented by a vector $\epsilon$ with elements $\epsilon_i \in \{-1,1\}$, such that $\epsilon$ is of length $n$ and each $\epsilon_i$ represents the binary state of neuron $i$. For $m$ stored memories $\epsilon^{(1:m)}$, the elements $w_{ij}$ of the network's weight matrix $w$ can be trained using Hebbian rules such that:
\begin{eqnarray}
w_{ij} = \frac{1}{m}\sum_{\mu=1}^m\epsilon_i^{\mu}\epsilon_j^{\mu}
\end{eqnarray}
where $\epsilon_{i}^{\mu}$ represents bit $i$ from pattern $\mu$ and $\epsilon_{j}^{\mu}$ represents bit $j$ from the same pattern. Each element $w_{ij}$ represents the interaction weight between neuron $i$ and neuron $j$. Note that $w_{ij}$ has the restriction $w_{ii}=0, \forall i$ such that no neuron has a connection with itself.

Let the vector $x$ describe the initial state of the Hopfield network with $n$ neurons where each neuron state is represented by $x_i \in \{-1,1\}$. The network undergoes a series of update steps where individual neurons have their states overwritten according to the interaction of the network state $x$ with the weight matrix $w_{ij}$. Neurons are updated using the rule:
\begin{eqnarray}
x_i=\begin{cases}
+1 &\sum\limits_{j=1}^{n}w_{ij}x_j>h_i\\
-1 &\text{else}
\end{cases}
\end{eqnarray}

\noindent where $h_i$ is the threshold of neuron $i$ and where each $i$ to be updated can be chosen randomly or in a predefined order. One can also define $\theta_i = \sum_{j=1}^{n}w_{ij}x_j$ in order to simplify the notation, where $\theta_i$ is known as the input signal to the neuron $i$. As successive updates occur, the state of the network will converge to a state which is a local minimum of the energy function $E=-\frac{1}{2}\sum_{i,j}w_{ij}x_ix_j + \sum_ih_ix_i$. Each of the stored memories are local minima of this energy function, and thus the stored memories are intuitively known as ``attractors" since the network state will converge to the attractor which is nearest in Hamming distance. The initial state of the network is also known as a ``probe," as it is used to search the associative memory and return the nearest attractor. In this work, we mimic these attractor dynamics in the quantum regime to create a Quantum Hopfield Associative Memory.

\subsection*{Background on Quantum Neuron Design}
\begin{figure}[!t]
\centering
\includegraphics[width=\columnwidth]{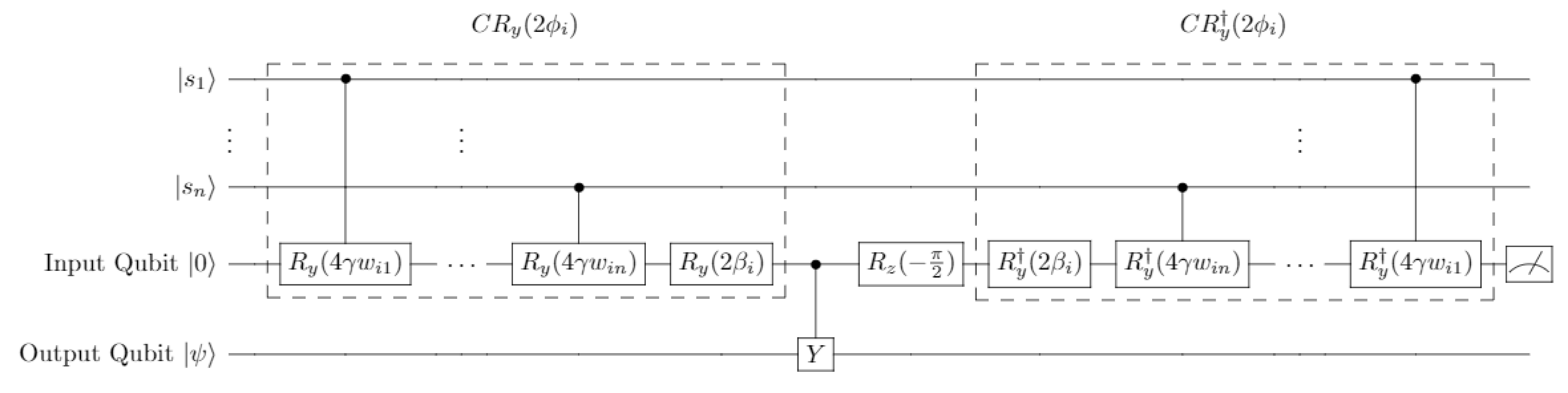}
\caption{\textbf{Quantum neuron circuit architecture based on Cao et al.\cite{GG_QN}} The basic quantum neuron architecture designed by Cao et al.\cite{GG_QN} utilizes controlled rotations dependent on a superposition of input states to create a controlled rotation $CR_y(2\phi)$. A measurement of $\ket{0}$ on the input qubit indicates that the output qubit has been rotated $R_y(2q(\phi))$ with $q(\phi)=arctan(tan^2\phi)$. If $\ket{1}$ is measured, the circuit has applied $R_y(\frac{\pi}{2})$ to the output qubit, in which case this rotation is reversed and the circuit is repeated until success is achieved.}
\label{Fig1}
\end{figure}

Several artificial neuron designs have been developed to apply this classical machine learning paradigm to quantum machine learning applications.\cite{GG_QN,NatureNeuron,QuantumPerceptron,ModifiedGG} We base our system on the quantum neuron developed by Cao et al. \cite{GG_QN} and modify their structure for implementation in IBMQ in a single, uninterrupted quantum circuit. The basic structure of the neuron developed by Cao et al. is shown in Fig. \ref{Fig1}.

In a quantum system, each classical neuron state $x_i$ can be mapped to a qubit in state $\ket{0}$ or $\ket{1}$. Mapping each neuron state to the probability of measuring $\ket{1}$ in the qubit, $x_i = -1$ corresponds to a pure $\ket{0}$ state ($P(\ket{1})=0$) and $x_i = 1$ corresponds to a pure $\ket{1}$ state ($P(\ket{1})=1$). Any non-classical value of $x_i\in (-1,1)$ can be represented as a superposition state, such as $x_i=0$ being represented by $\ket{s_i} = \frac{\ket{0}+\ket{1}}{\sqrt{2}}$ where $P(\ket{1})=0.5$.\cite{GG_QN} One possible mapping of classical states to quantum states in this manner is described by:
\begin{eqnarray}
\ket{s_i} = cos(x_i\frac{\pi}{4}+\frac{\pi}{4})\ket{0}+sin(x_i\frac{\pi}{4}+\frac{\pi}{4})\ket{1}
\end{eqnarray}

The full quantum system is thus represented by $\ket{s}=\ket{s_1s_2 \cdots s_n}$. In order to mimic the attractor dynamics of a classical Hopfield network, we desire an update rule similar to equation (2) which can update the state of a single quantum neuron $\ket{s_i}$ based on the interaction of $w_{ij}$ and $\ket{s}$ compared to a given threshold. In order to accomplish this, Cao et al.\cite{GG_QN} consider the quantum circuit shown in Fig. \ref{Fig1} which produces a rotation on a target qubit of $R_y(2arctan(tan^2\phi))$ if the input qubit measures $\ket{0}$, where $R_y$ represents a rotation generated by the Pauli Y operator around the Y-axis of the Bloch sphere. This rotation, when applied to an initial $\ket{0}$ state, produces a $\ket{1}$ measurement probability of $P(\ket{1})=\frac{sin^4(\phi)}{sin^4(\phi)+cos^4(\phi)}$ on the output qubit. This output probability is used as an analog to the classical neuron, in which $\phi > \frac{\pi}{4}$ produces $P(\ket{1})$ close to $1$ and $\phi < \frac{\pi}{4}$ produces a $P(\ket{1})$ close to $0$, corresponding to the classical $x_i =\pm1$ states. The parameter $\phi$ relates to the classical $\theta$ parameter by:
\begin{eqnarray}
\phi = \gamma\theta + \frac{\pi}{4}
\end{eqnarray}
where $\gamma$ is a normalization factor such that $0<\phi<\frac{\pi}{2}$. Using this construction, the threshold behavior is achieved following equation (2) where all $h_i=0$. The function chosen for $\gamma$ is:
\begin{eqnarray}
\gamma=\frac{\frac{\pi}{4}}{w_{max}n}
\end{eqnarray}
We have slightly modified $\gamma$ from the Cao et al. design\cite{GG_QN} by removing the threshold term since all $h_i=0$ in the Hopfield associative memory application, and by using $\frac{\pi}{4}$ in the numerator instead of the approximate value of $0.7$. This helps to ensure that quantum states corresponding to the classical $x_i = 1$ are as close as possible to $\ket{s_i} = \ket{1}$.

The $CR_y(2\phi_i)$ rotation gate in the circuit in Fig. \ref{Fig1} is composed of rotations of the neuron state $\ket{s_i}$ controlled by each $\ket{s_j}$ which depend on the neural interconnection matrix elements $w_{ij}$, followed by a rotation dependent on the bias term $\beta$. The bias term $\beta$ is set by:
\begin{eqnarray}
\beta_i=\frac{\pi}{4}-\gamma(\sum_{j=1}^{n}w_{ij})
\end{eqnarray}
which we simplify from the original design\cite{GG_QN} using the knowledge that all $h_i=0$. When these rotations are combined, the resulting operation is a rotation $R_y(2\phi_i)$ on the target qubit.

This circuit operates in a Repeat-Until-Success (RUS)\cite{RUS} manner wherein a measurement of $\ket{0}$ on the input qubit indicates that the rotation of $R_y(2arctan(tan^2\phi))$ has been successfully applied to the output qubit. If the input qubit measures $\ket{1}$, this indicates that a rotation of $R_y(\frac{\pi}{2})$ has been applied instead, in which case $R_y(-\frac{\pi}{2})$ is applied to reverse the initial rotation and the circuit is repeated. This process continues until $\ket{0}$ is measured on the input qubit to indicate a successful rotation of the output qubit by $R_y(2arctan(tan^2\phi))$. Additionally, Cao et al. introduce a recursion method for this circuit which sharpens the rotation function to $R_y(2(arctan(tan^2\phi))^{\circ k})$, where $k$ is the number of recursions. This function performs more closely to a binary activation function as in equation (2) and thereby reduces the error rate of the system. Throughout this work, we compare our system to the $k=1$ case of this design which does not include recursion.

This quantum neuron design is used to model Hopfield attractor dynamics wherein $\phi>\frac{\pi}{4}$ (equivalently, $\theta>h=0$) results in $P(\ket{1})$ close to $1$ and $\phi<\frac{\pi}{4}$ results in $P(\ket{1})$ close to $0$. The derivation and characterization of this quantum neuron model is shown and described in more detail by Cao et al.\cite{GG_QN}

\subsection*{Challenges in Hardware Implementation}

The RUS circuit design presents difficulty for implementation in quantum hardware due to its requirement of mid-circuit measurement and reset operations. The mid-circuit measurements are used to perform the check for success and the reset operations are used to recycle the output qubit for successive updates. These operations are a limiting factor for the hardware implementation of many quantum algorithms, as measuring or resetting the state of a qubit can result in collapsing the wavefunction of other qubits in the system. IBMQ is one quantum development platform which has recently enabled reset operations in its newest hardware for dynamic circuit execution,\cite{dynamicQC} but avoiding qubit reset in favor of sequential circuit execution remains desirable for many quantum hardware platforms and applications.

The qubit overhead of the Cao et al. design is dependent on the use of reset operations in order to reset the ancilla qubits (the input and output qubits in Fig. \ref{Fig1}) to $\ket{0}$ to recycle them for many qubit updates.\cite{GG_QN} Should the system be implemented in hardware where reset is not available, this overhead increases from $n+2$ qubits (where $n$ is the size of the associative memory) to $n+2u$ qubits, where $u$ represents the number of updates to be performed on the system. This qubit overhead is included in Table 1.

In its basic form, the Cao et al. quantum neuron design \cite{GG_QN} also has very high gate complexity, thereby incurring extensive time overhead upon execution. The gate complexity of this design is shown in Table 1 and derived in more detail in the Supplementary Information. The high gate complexity of the quantum neuron becomes a significant problem for the overall error rate of the system. As errors occur in each gate, the large number of gates leads to significant error in the overall output of the system as errors cascade from one gate operation to the next. A metric called quantum volume \cite{quantum_volume} describes the largest circuit of equal width and depth that a quantum computer can successfully implement. Considering the small quantum volume of modern quantum systems in the NISQ era (shown for IBMQ systems in Supplementary Information Table S1), hardware execution of a circuit as complex as an associative memory requires low circuit complexity to be successfully implementable in hardware. Reducing gate complexity thus becomes a very important factor to the design of a QHAM.

\begin{figure}[!t]
\centering
\includegraphics[width=0.6\columnwidth]{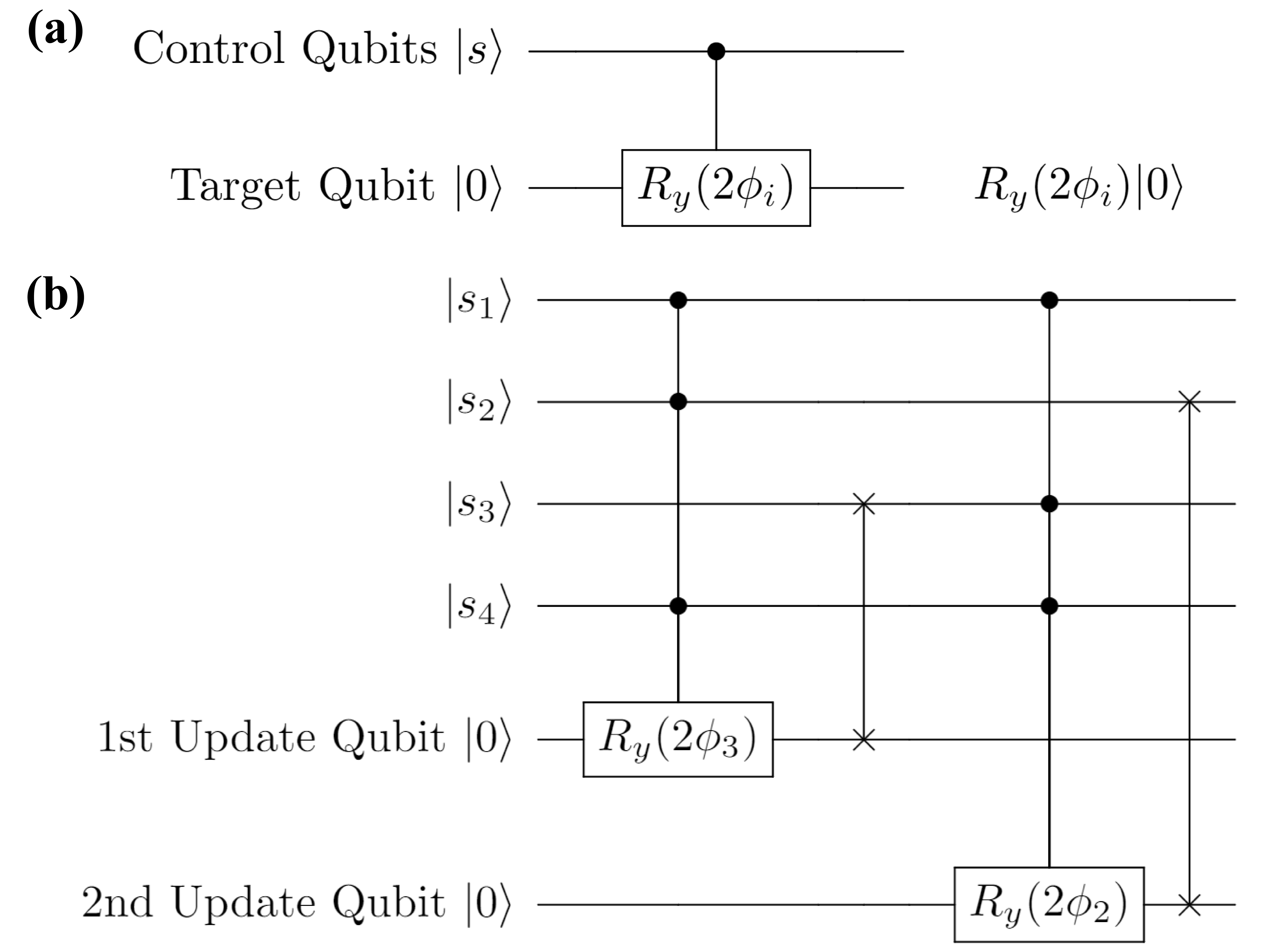}
\caption{\textbf{Proposed quantum neuron circuit structure. (a)} Our proposed quantum neuron for quantum hardware implementation. The full controlled rotation $CR_y(\phi_i)$ breaks down into sub-rotations dependent on each control qubit as shown in Fig. \ref{Fig1}. \textbf{(b)} Quantum circuit representing two updates of a selected qubits, $\ket{s_3}$ then $\ket{s_2}$. In this case, $\ket{s_3}$ is updated using the state described by $\ket{s_1 s_2 s_4}$, then $\ket{s_2}$ is being updated using the state described by $\ket{s_1 s_3' s_4}$.}
\label{Fig2}
\end{figure}

\subsection*{Proposed Quantum Neuron}

To address the challenges in the original quantum neuron design,\cite{GG_QN} we propose a quantum neuron design which avoids mid-circuit measurement and reset operations and which improves resilience to errors in quantum hardware. In order to avoid these operations, the measurement operation on the input qubit in the neuron described in Fig. \ref{Fig1} is removed. In the design by Cao et al.,\cite{GG_QN} this measurement was used for classical conditioning to determine whether or not the circuit construction must be repeated in the Repeat-Until-Success construction. By removing this conditioning step, the probability of measuring $\ket{1}$ at the output qubit is changed from $\frac{sin^4(\phi)}{sin^4(\phi)+cos^4(\phi)}$ to simply $sin^2(\phi)$, which is the same output probability as that which is obtained by applying a rotation of $CR_y(2\phi)$. Therefore, we modify our quantum neuron from the original structure in Fig. \ref{Fig1} to simply be a $CR_y(2\phi)$ rotation controlled by all other qubits in the system, as shown in Fig. \ref{Fig2} and following the same construction as the boxed $CR_y(2\phi_i)$ rotation in Fig. \ref{Fig1}. 

This reduction removes mid-circuit measurement operations and significantly lowers the circuit complexity of the neuron design, as only a single multi-controlled rotation is performed rather than the full, repeated construction shown in Fig. \ref{Fig1}. A summary of the differences between our neuron and the neuron by Cao et al.\cite{GG_QN} is shown in Table 1. As derived in the Supplementary Information, the gate complexity of our design is only $(10n-3)u$, where $n$ is the size of the associative memory and $u$ is the number of updates performed. By comparison, the gate complexity of the RUS neuron is $[20n(f+1)-4f-5]u$, where $f$ is the average number of failures in the RUS process. This significant reduction in gate complexity reduces the time overhead of the circuit and reduces the effect of cascading errors in each qubit operation. Additionally, our design can be rid of reset operations by not recycling the ancilla (target) qubit for successive updates at the cost of qubit overhead.

To update an initial qubit state $\ket{s_i}$ to a new state $\ket{s_i}'$, the rotation by $2\phi$ is performed on an ancilla (target) qubit initialized to state $\ket{0}$, then the output of the neuron on the ancilla qubit is swapped to the qubit which held the original state using a quantum SWAP gate. An example of this circuit using an initial state of length $n=4$ and performing an update of the third qubit $\ket{s_3}$ followed by an update of the second qubit $\ket{s_2}$ is shown in Fig. \ref{Fig2}b. In this example, the qubit used to perform the update is not recycled in order avoid reset operations. Avoiding reset operations improves the accuracy of the neuron and allows for implementation in hardware which does not support reset operations. In order to save on qubit overhead in systems where reset is possible, one can recycle the ancilla for multiple updates by performing the $CR_y$ rotation, swapping the updated state to the desired qubit, resetting the ancilla to $\ket{0}$, and performing the next update. Since the quantum neuron design requires the target qubit to be in state $\ket{0}$ before performing the rotations, one cannot simply perform the rotation on the desired qubit directly. However, the SWAP operation can also be avoided by reassigning the target qubit to be the $\ket{s_i}'$ and discarding the qubit which originally held $\ket{s_i}$.

In the original neuron,\cite{GG_QN} a majority vote of many measurements of the updated qubit is used to reassign the state to pure $\ket{0}$ or $\ket{1}$ immediately after being updated in order to achieve a binary activation function. As this requires mid-circuit measurement and repeated execution of the circuit, we leave the updated qubits in their state produced by the update between evaluation steps as shown in Fig. \ref{Fig2}b. Since this means that the update step does not produce a true binary result reassigning qubit states to true $\ket{0}$ and $\ket{1}$, there will be an inherent error rate as a result of the update step with respect to the true classical Hopfield update. Multiple updates are performed in succession to reach a desired endpoint of the circuit, after which the majority vote method can used to extract the result to a classical output, or further evaluation can be continued in the quantum domain.

\begin{figure}[!t]
\centering
\includegraphics[width=0.9\columnwidth]{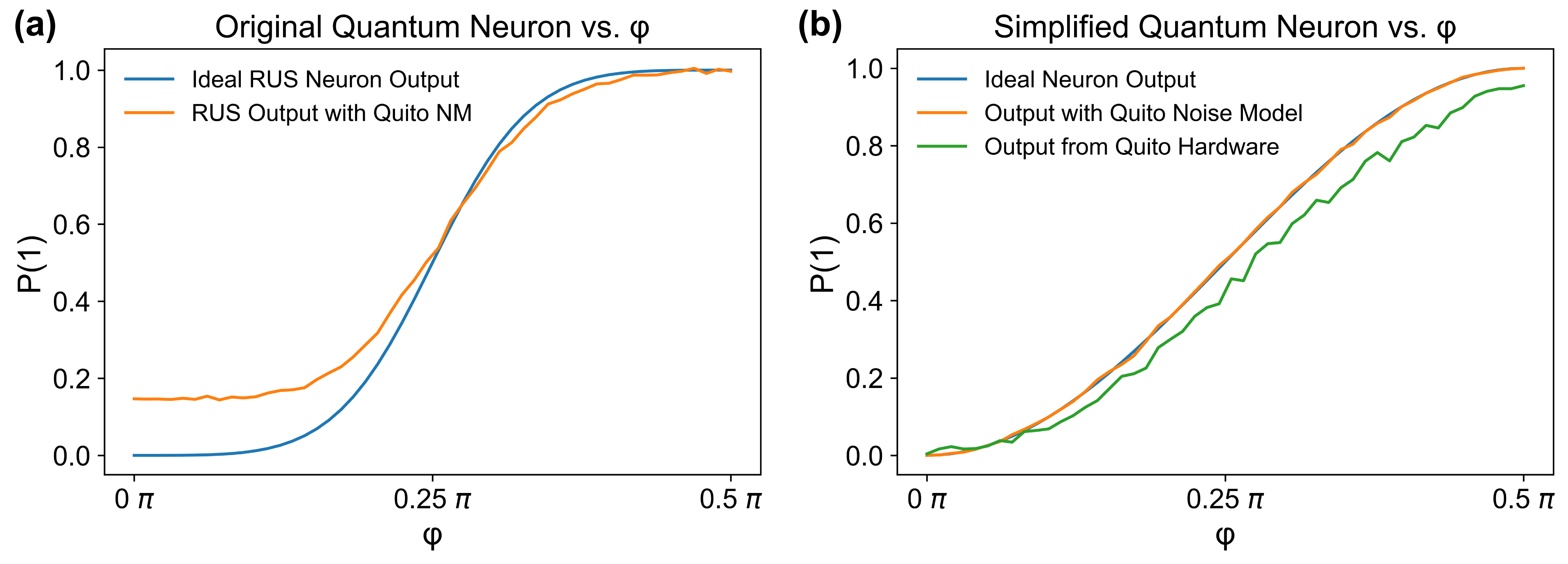}
\caption{\textbf{Quantum neuron output.} \textbf{(a)} Simulation of the output of the neuron designed by Cao et al. \cite{GG_QN} based on Repeat-Until-Success (RUS) operation using the \textit{ibmq\_quito} \cite{IBMQ_qasm,ibmq_quito} noise model. \textbf{(b)} Simulation and IBMQ hardware \cite{IBMQ_qasm,ibmq_quito} results of our simplified quantum neuron compared to the ideal output. }
\label{Fig3}
\end{figure}

\begin{table*}[!t]
\caption{\textbf{A summary of the differences between our quantum neuron and the quantum neuron designed by Cao et al.\cite{GG_QN}} In these calculations, it is assumed that reset is unavailable and there is no recursion in the Cao et al. neuron. The parameter $n$ represents the size of the associative memory, $u$ represents the number of updates, and $f$ represents the average number of failures in the RUS process. See the Supplementary Information for a more detailed analysis and derivation of gate complexity.}
\centering
\label{Table1}
\begin{threeparttable}
\begin{center}
\begin{tabular}{ | c | c | c | }
 \hline
 Neuron Characteristic & RUS Neuron Design \cite{GG_QN} & Our Neuron Design \\ 
 \hline
 Output $P(\ket{1})$ & $\frac{sin^4(\phi)}{sin^4(\phi)+cos^4(\phi)}$ & $sin^2(\phi)$ \\
 \hline
 Mid-Circuit Measurement Required & Yes & No \\ 
 \hline
 Qubit Overhead Without Resets & $n+2u$ & $n+u$ \\ 
 \hline
 Qubit Overhead With Resets & $n+2$ & $n+1$ \\ 
 \hline
 Total Gate Complexity & $O(nuf)$ & $O(nu)$ \\ 
 \hline
 Single-Qubit Depth & $(16n(f+1)-f-5)u$ & $(8n-4)u$ \\ 
 \hline
 CNOT Depth & $(4n(f+1)-3f)u$ & $(2n+1)u$ \\ 
 \hline
 Majority Vote & After Each Update & After All Updates \\
 \hline
\end{tabular}
\begin{tablenotes}
      \small
      \item
\end{tablenotes}
\end{center}
\end{threeparttable}
\end{table*}

The output of the RUS neuron is shown in Fig. \ref{Fig3}a. Using a noise model defined by IBMQ which models \textit{ibmq\_quito} hardware noise, we observe a significant error rate particularly in the region $\phi<\frac{\pi}{4}$. This neuron cannot be tested in the \textit{ibmq\_quito} hardware because of the required mid-circuit measurement and reset operations and because of the limited size of the \textit{ibmq\_quito} architecture. We expect that this quantum neuron would perform significantly worse in true hardware than in noise model simulation considering the connectivity constraints in hardware mentioned leading to increased gate complexity.

The simulation and hardware accuracy of our simplified quantum neuron is shown in Fig. \ref{Fig3}b. We observe that our updated design is more resilient to noise and hardware errors than the RUS design because it requires less gates and less qubits. The simulation results of our neuron design incorporating the \textit{ibmq\_quito} noise model shown in Fig. \ref{Fig3}b almost exactly overlap with the desired output function. The hardware measurements show degradation particularly in the $\phi>\frac{\pi}{4}$ regime, but this hardware error is much lower than the simulated error in the RUS design.

In the original design, an activation function was applied to the rotation angle of the output state, namely $arctan(tan^2(\phi))$, whereas in our design, we allow for the rotation around the Bloch sphere itself to act as an activation function, such that the probability of measuring $\ket{1}$ follows the non-linear function $P(\ket{1})=sin^2(\phi)$ dependent on $\phi$. This reduction results in a more gradual activation function between measuring $\ket{0}$ and $\ket{1}$ (as shown in Fig. \ref{Fig3}) but allows for uninterrupted execution in a single quantum circuit and reduces the qubit overhead and gate complexity of the circuit. While the slope of the activation function in our neuron is more gradual, it also shows much better resilience to noise and demonstrates performance in hardware which closely models its expected output, as shown in Fig. \ref{Fig3}. However, this more gradual slope of the activation function can reduce the accuracy of update steps where $\phi$ is close to its midpoint of $\frac{\pi}{4}$.

\subsection*{The Quantum Hopfield Associative Memory}

To model an associative memory using our quantum neuron, an initial state vector $\bar{x}$ is given and converted to its quantum representation $\ket{s}=\ket{s_1 s_2 \cdots s_n}$. Given this initial state and a set of attractor states trained into the weight matrix $w$ as described by equation (1), updates are performed on individually selected qubits to move the network state closer to the nearest attractor. The qubit overhead of the system with attractors and initial state of length $n$, which we define as $q$, is determined by:
\begin{eqnarray}
q = n+1
\end{eqnarray}
given that the test is performed in an environment where the qubit used for the updates can be recycled using a reset operation. However, when qubits cannot be recycled in hardware, an additional qubit is required per update (as in Fig. \ref{Fig2}). When performing $u$ updates, equation (7) becomes:
\begin{eqnarray}
q = n+u
\end{eqnarray}
This qubit overhead is included in Table 1.

Note that there are fundamental differences in the way updates are performed in our design and in the classical Hopfield associative memory. In the classical case, each update results in the updated $x_i$ being rewritten to $-1$ or $+1$ depending on the value of the update rule with respect to the desired threshold, typically described by equation (2). However, in our quantum case, we utilize a rotation around the Bloch sphere where $P(\ket{1})=sin^2(\phi_i)$ and therefore the updated qubit states are rarely a true $\ket{0}$ or $\ket{1}$. Cascading many updates on all qubits in the system in this fashion without stopping execution of the quantum circuit intermediately to rewrite the qubit states to pure $\ket{0}$ or $\ket{1}$ based on a majority vote measurement value (as in Cao et al.\cite{GG_QN}) will eventually result in the state of each qubit converging to a superposition state depending on the average value of the probe elements rather than converging to a memory state as desired. Therefore, in order to maintain the quantum nature of the system within the capabilities of quantum hardware, we perform either individually selected updates on specific qubits or random updates up to a set hyperparameter of $u$ total updates which can be tuned for maximum accuracy. After all of the updates, the system state can be measured with a desired number of ``shots'' to determine the probability distribution for each state. To determine the classical output of the system, we take a majority vote of many measurements of each of the individual qubits in the system to compose the final memory recall result.

\subsubsection*{Simulating the QHAM}
\begin{figure}[t]
\centering
\includegraphics[width=\columnwidth]{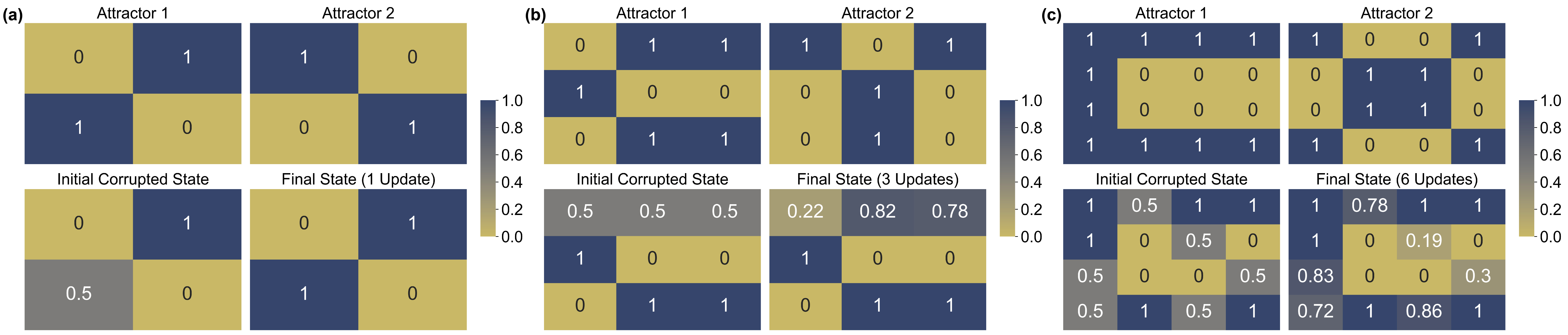}
\caption{\textbf{QASM simulation of QHAM of size 4, 9, and 16.} Simulation results of the quantum Hopfield associative memory system for sizes \textbf{(a)} 4, \textbf{(b)} 9 and \textbf{(c)} 16 \cite{IBMQ_qasm} with two given attractor states. Each color graded box represents a qubit in each system state, and the number inside the box is the exact probability of measuring $\ket{1}$ for that individual state.}
\label{Fig4}
\end{figure}

\begin{figure}[!t]
\centering
\includegraphics[width=0.65\columnwidth]{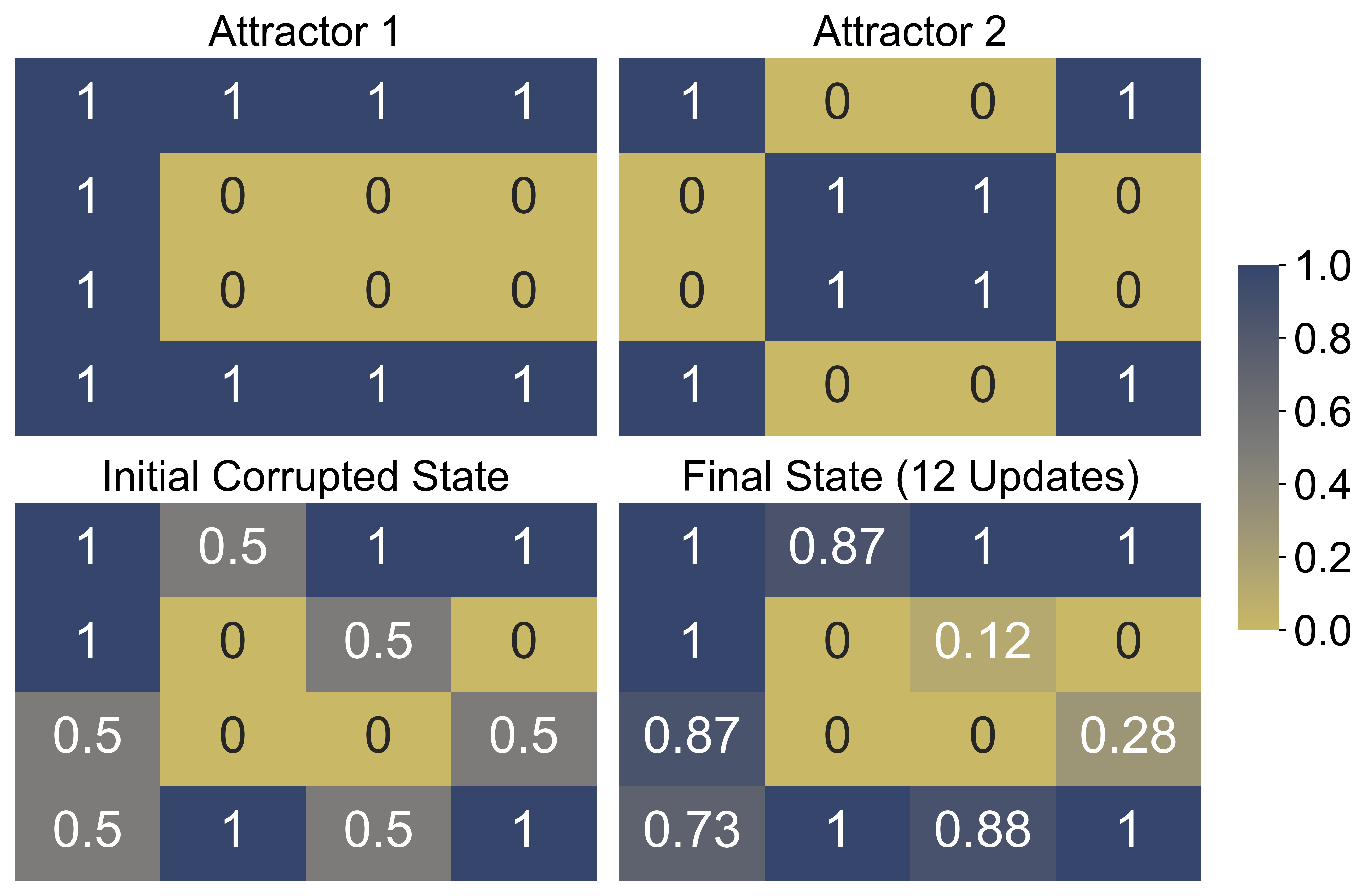}
\caption{\textbf{QHAM multiple update demonstration.} Updating each corrupted qubit from the Fig. \ref{Fig4}c memory twice improves the accuracy of the several of the updated qubits, particularly those which are towards the beginning of the update order.}
\label{Fig5}
\end{figure}

Simulation results for two-attractor systems of sizes 4, 9 and 16 are shown in Fig. \ref{Fig4}. The vectors representing the system states are reshaped to squares where the index of the memory elements are read left to right, top to bottom. In the size 9 and 16 cases, we update multiple selected qubits which are in ``corrupted'' superposition states of $\frac{\ket{0}+\ket{1}}{\sqrt{2}}$ (equivalent to $x_i=0$ described previously), starting from lowest index corrupted qubit and following in order of index to demonstrate the performance of individual updates with our quantum neuron. Qubits which are updated first in this order tend to have a lower accuracy than qubits which are updated later in the update chain, since in the early updates there are more corrupted states used as controls for the update. This can be alleviated by updating the qubits multiple times, as in Fig. \ref{Fig5}. Note that Figs. \ref{Fig4} and \ref{Fig5} show the $P(\ket{1})$ of each qubit. From this, a majority vote of each of the qubit measurements can be taken to compose a classical binary result. One potential application of this execution method is in restoring images with corrupted pixels.

\subsubsection*{Testing the QHAM in IBMQ Hardware}
\begin{figure}[!t]
\centering
\includegraphics[width=0.65\columnwidth]{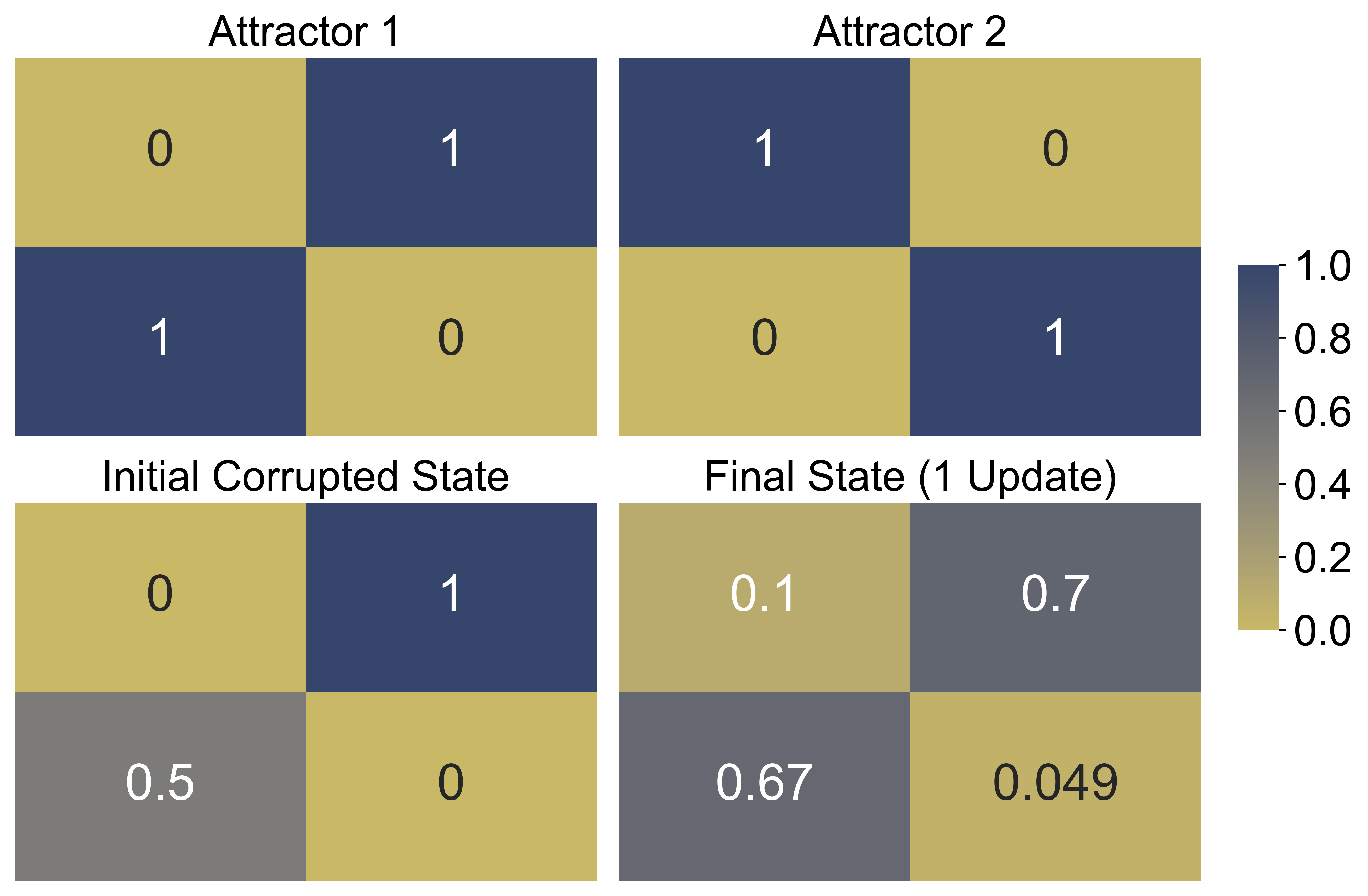}
\caption{\textbf{QHAM implemented in quantum hardware.} Hardware results for the $n=4$, single update system obtained from the \textit{ibmq\_16\_melbourne} device.\cite{ibmq_16_melbourne} The numbers in each box (representing each qubit) represent the probability of measuring $\ket{1}$ for each qubit. Two attractors are given and an initial state with $\ket{x_3}$ corrupted to a superposition state is updated to the final state.}
\label{Fig6}
\end{figure}

We also demonstrate the $n=4$ system in the 15-qubit \textit{ibmq\_16\_melbourne} hardware \cite{ibmq_16_melbourne} shown in further detail in Fig. S1 of the Supplementary Information. The hardware results for the $n=4$ system execution are shown in Fig. \ref{Fig6}. The ideal $n=4$ system with a single update should require 5 qubits for its execution according to equation (7). However, since each qubit in the device is not directly physically connected to all other qubits, intermediate qubits must be used in order to transfer information between control and target qubits. As a Hopfield network requires high connectivity, the overall qubit connectivity of the system is crucial for performance in order to avoid these intermediate connections and additional qubits which contribute to increased noise in the system. Further analysis of hardware limitations and noise can be found in the Supplementary Information, including Table S1 and Table S2.

\subsection*{Determining Effective Memory Capacity}
To benchmark the QHAM we study its effective memory capacity \cite{McEliece_Capacity,Newman_Capacity} which quantifies the number of memories $m$ which can be successfully stored and recalled in an associative memory of size $n$. We test the effective memory capacity using a similar method to the classical Hopfield associative memory capacity calculations developed by McEliece et al. \cite{McEliece_Capacity} and Newman.\cite{Newman_Capacity}

First, we introduce $m$ patterns of length $n$ which are composed of random, independent values of $\pm1$. These patterns are used as attractors with the set of all attractors described by $\epsilon^{(1:m)}$ and are used to develop the weight matrix according to equation (1). We then introduce a random probe (input state) of length $n$ which is less than $\rho n$ in Hamming distance from one of the stored memory patterns. Each memory or probe can be described by $x=[x_1,x_2,\cdots,x_n]$ with $x_j\in\{-1,1\}$. For example, with $m=2$, $n=5$ and $\rho=0.25$, we could consider $\epsilon_1 = [1,1,1,1,1]$ and $\epsilon_2 = [-1,-1,-1,-1,-1]$. An acceptable probe within $\rho n = 0.25n$ from one of these attractors could have at most one flipped bit from its closest attractor, i.e. the probe could be $x = [-1,1,1,1,1]$ which is within $\rho n$ of $\epsilon_1$, or $x = [-1,-1,1,-1,-1]$ which is within $\rho n$ of $\epsilon_2$. As the Hopfield network performs updates on random elements of the probe, the state of the probe converges to the stored memory which is the closest in Hamming distance to the original probe. In a rigorous analysis by McEliece et al.,\cite{McEliece_Capacity} it is derived that with $0\leq\rho<\frac{1}{2}$, the maximum $m$ patterns which can be successfully recalled in a classical associative memory grows as:
\begin{eqnarray}
m=\frac{(1-2\rho)^2}{2}\frac{n}{ln(n)}
\end{eqnarray}

We map these $m$ memories and the probe to the quantum domain following equation (3), prepare the input state described by the equivalent quantum representation of the probe $s=\ket{s_1s_2\cdots s_n}$ and create the Hebbian weight matrix using the attractors following equation (1). Random updates are performed up to the hyperparameter $u$ (representing the number of updates to perform) and the output of all $n$ states are measured. After all of the updates, the system state can be measured with a desired number of ``shots'' to determine the probability distribution for each state. To determine the classical output of the system, we take a majority vote of many measurements of each of the individual qubits in the system to compose the final memory recall result.

\begin{figure*}[!t]
\includegraphics[width=\textwidth]{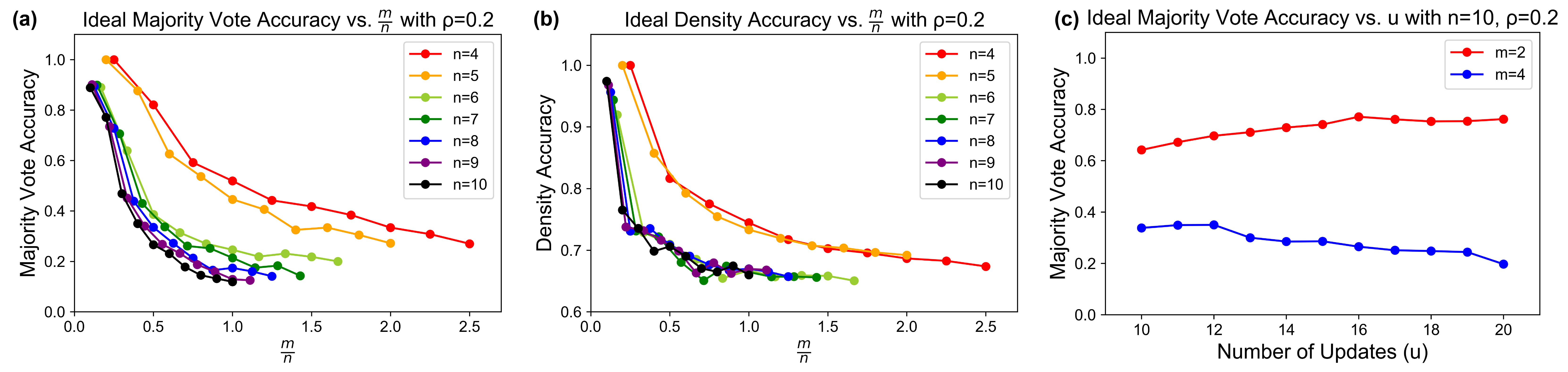}
\caption{\textbf{Effective QHAM Capacity. (a)} Majority vote accuracy measurements for the QHAM for memories of length $n\in\{4,5,\cdots,10\}$ and $m\in\{1,2,\cdots,10\}$ vs $\alpha=m/n$. \textbf{(b)} Density accuracy measurements from the same test conditions as (a) vs $m/n$. \textbf{(c)} Demonstration of hyperparameter tuning for $u$. In this case, the optimal $u$ based on majority vote accuracy is 16 for $m=2$ and 12 for $m=4$.}
\label{Fig7}
\end{figure*}

\begin{figure}[t]
\includegraphics[width=\columnwidth]{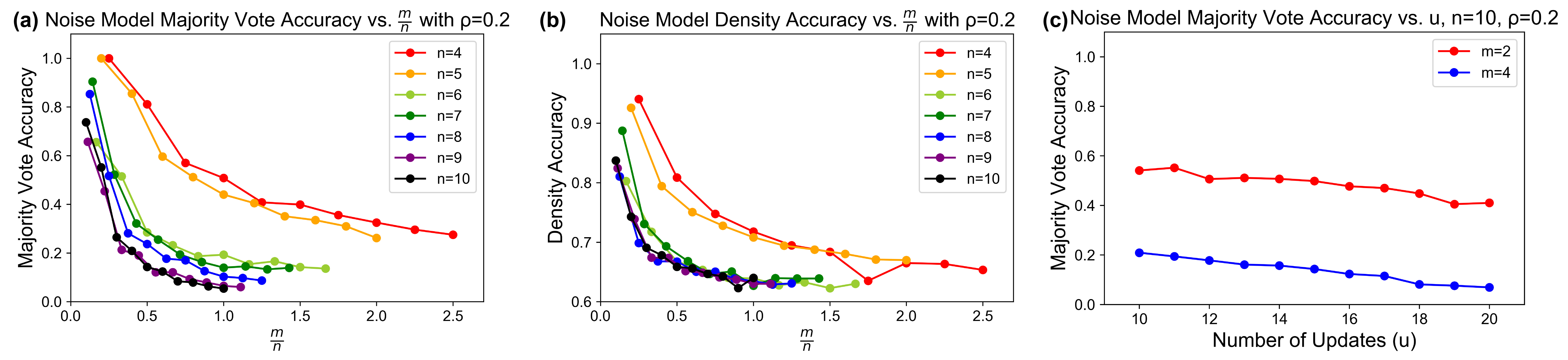}
\caption{\textbf{Effective QHAM Capacity using \textit{ibmq\_16\_melbourne} noise model. (a)} Majority vote accuracy for the QHAM for memories of length $n\in\{4,5,\cdots,10\}$ and $m\in\{1,2,\cdots,10\}$ vs $\alpha=\frac{m}{n}$. \textbf{(b)} Density accuracy from the same test conditions as (a) vs $\frac{m}{n}$. \textbf{(c)} Demonstration of tuning for $u$. In this case, the optimal $u$ based on majority vote accuracy is 11 for $m=2$ and 10 for $m=4$.}
\label{Fig8}
\end{figure}

\subsubsection*{Simulation of QHAM Effective Memory Capacity}
To analyze effective memory capacity, we perform Monte Carlo simulation with many random memories and probes to determine the accuracy of the system in reproducing the desired memory state. The capacity results for $n=\{4,5,\cdots,10\}$, $m=\{1,2,\cdots,10\}$ and $\rho=0.2$ are shown in Fig. \ref{Fig7}. Figure \ref{Fig7}a shows these results as a function of $\alpha=\frac{m}{n}$ using a term we call ``majority vote accuracy,'' which is measured as the percentage of accurate memory strings obtained after taking a majority vote of many measurements of each of the individual qubits in the system. For example, if an experiment is run with 8192 ``shots'' as we use for our simulations, we take a majority vote for each qubit such that if a majority of the 8192 shots return $\ket{s_0}=\ket{0}$, then result of qubit $\ket{s_0}$ is assigned to $\ket{0}$. For a target memory of $\ket{s}=\ket{0110}$, the majority vote accuracy is measured as the percentage of tests in which $\ket{0110}$ is returned after the majority vote result of each individual qubit is taken.

In Fig. \ref{Fig7}b, we analyze an accuracy metric which we call ``density accuracy'' which describes the average rate of each of the individual qubits being measured in the correct state. This measurement quantifies the average accuracy of each individual qubit in the system and is used to model how the results propagate in the quantum domain, as it is a close reflection of the superposition states of each of the individual qubits. For example, if $\ket{s_0}$ returns the desired value $\ket{0}$ in $90\%$ of shots, and $\ket{s_1}$ returns the desired value $\ket{1}$ in $100\%$ of shots, then the density accuracy of the two qubit system is $95\%$. The results in Fig. \ref{Fig7}a and \ref{Fig7}b have been tuned to the value of the hyperparameter $u$ which yields the maximum majority vote accuracy. An example of this hyperparameter tuning is shown in Fig. \ref{Fig7}c. Since we cannot check for convergence in the quantum domain after each individual update because this would require mid-circuit measurement, we tune $u$ with Monte Carlo simulation.

In these QHAM implementations, we notice trends mirroring the classical memory capacity calculations, even with the neuron activation function not being perfectly binary (as shown in Fig. \ref{Fig3}). With $\rho=0.2$ and $n\in [4,5,\cdots,10]$, the probes which are allowed to be introduced contain no flipped bits when $n=4$ or $n=5$, and a single flipped bit for $6\leq n \leq 10$. Based on these results, we clearly see two separate trends in Fig. \ref{Fig7} wherein a greater number of flipped bits in the probe leads to lower classification accuracy. Since the $\rho$ we define is an upper bound, we can also calculate an effective $\rho$, $\rho_{eff}$, which corresponds to the true percentage of bits in the probe which are flipped. For the $n=4$ and $n=5$ cases this $\rho_{eff}=0$, and for the case of $n=10$, for example, $\rho_{eff}=0.1$. Following equation (9) using $\rho_{eff}$, we would then expect $m$ to be $1.44$, $1.55$, and $1.39$ for $n=4$, $5$ and $10$ respectively, which can be rounded down to the largest number of memories which can be stored for successful recall, specifically $m=1$ in these cases. With $\rho=0.2$ and $n\in [4,5,\cdots,10]$, we see the $m=1$ case for $n=4$ and $n=5$ demonstrate perfect recall in the majority vote accuracy test (which extracts the associative memory to the classical domain), and the $m=1$ cases for $n>5$ show very high recall percentage which increases with $n$. In the density accuracy measurements in Fig. \ref{Fig7}b, which reflect the accuracy of the system in the quantum domain, we observe well over $90$ $\%$ accuracy for all $n$ when $m=1$. We also demonstrate tuning of the $u$ parameter in Fig. \ref{Fig7}c based on the majority vote accuracy measurements. Fig. \ref{Fig7}c shows the results of tuning the QHAM of length $n=10$ when storing $m=2$ and $m=4$ memories. In these examples, the majority vote accuracy increases with each update as incorrect bits are corrected up to an optimal number of updates $u$. After this optimal $u$, the accuracy tends to decrease as the inaccuracy inherent in the neuron activation function (shown in Fig. \ref{Fig3}) propagates through the circuit and leads to the incorrect changing of correct bits in the memory string. With a larger number $m$ of stored memories, spurious minima are more likely to exist in the optimization landscape of the associative memory system, lowering the optimal $u$ and the overall accuracy of the system. Overall, the accuracy improves with increasing updates up until the optimal $u$, beyond which the probability of incorrectly flipping an incorrect bit outweighs the probability of correctly flipping an incorrect bit. In this example, the optimal $u$ is shown to be 16 for $m=2$ and 12 for $m=4$. All of the data shown in Figs. \ref{Fig7}a and \ref{Fig7}b are tuned using this metric.

\subsubsection*{Simulation of QHAM Effective Memory Capacity Considering Noise}
In addition to simulating the effective memory capacity of the QHAM system in the noiseless scenario, we perform measurements of the QHAM capacity under the effects of simulated hardware noise from the \textit{ibmq\_16\_melbourne} noise model as shown in Fig. \ref{Fig8}. The sources of this noise due to quantum errors are further described in the Supplementary Information. With $n=\{4,5,\cdots,10\}$, $m=\{1,2,\cdots,10\}$ and $\rho=0.2$ as in the noiseless simulations, we see the $m=1$ case for $n=4$ and $n=5$ demonstrate perfect recall when the output is extracted to the classical domain in the majority vote accuracy test in Fig. \ref{Fig8}a, and degradation of the recall for larger $n$. In these measurements, we observe that memories with larger size $n$ show lower effective capacity for the same $\rho_{eff}$. This is explained by the larger number of updates performed in these cases and the need to take majority votes from $n$ probabilistic qubit outputs wherein inaccuracies caused by noise can compound for large $n$. Without this inaccuracy from the majority vote extraction to the classical domain, the density accuracy metric shown in Fig. \ref{Fig8}b tends to show a more condensed trend regardless of $n$. However, compared to the same tests without any noise modeling as shown in Fig. \ref{Fig7}, the noise model results show significant degradation in their density accuracy measurements, for example the $m=1$, $n=4$ case showing approximately $95~\%$ accuracy as opposed to perfect accuracy in the noiseless test. We also demonstrate tuning of the $u$ parameter in Fig. \ref{Fig8}c based on the majority vote measurements. Tuning $u$ shows less of a defined optimization point when noise effects are included compared to the noiseless case, as increasing the number of updates leads to more compounding error due to noise. Increasing the number of updates increases the gate complexity of the quantum circuit, thereby increasing the amount of noise injected by each gate operation, thus increasing the error inherent in the neuron (see Fig. \ref{Fig3}). Therefore, the optimal $u$ tends to be lower in the noisy associative memory tests than in the equivalent noiseless associative memory, as the noise introduced by further gate complexity eventually outweighs the benefit of performing an update.

\begin{figure}[!t]
\centering
\includegraphics[width=0.8\columnwidth]{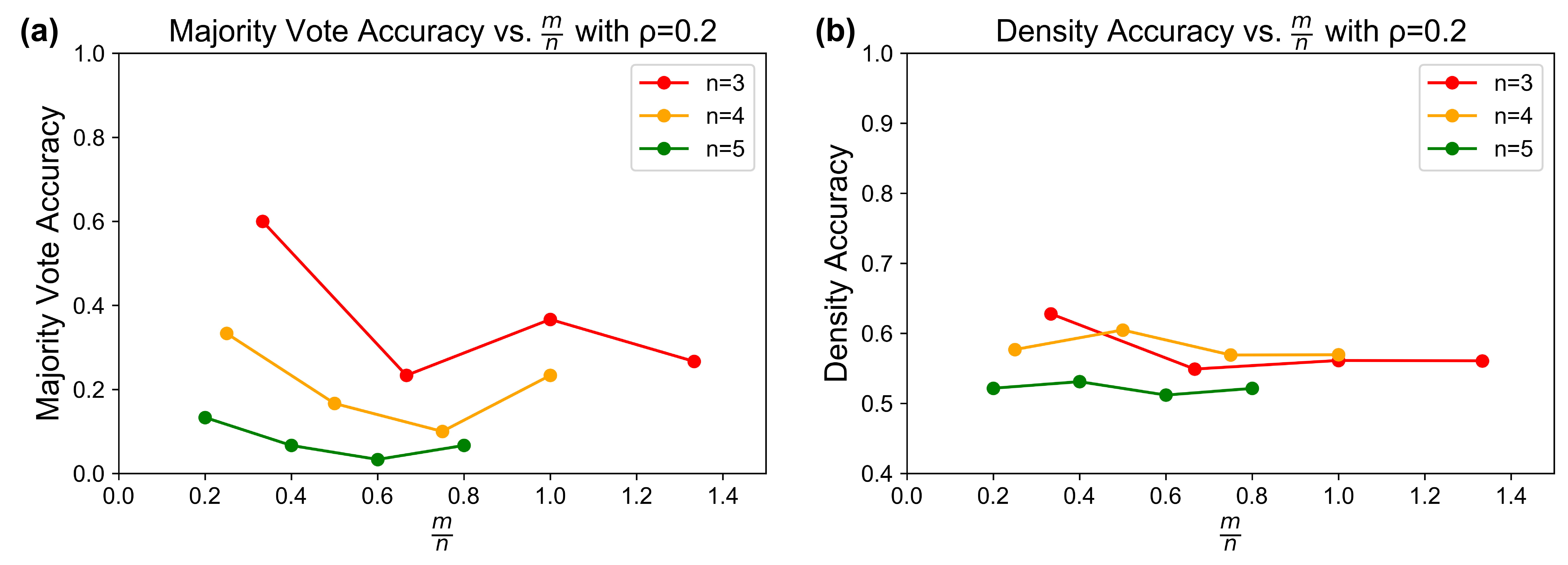}
\caption{\textbf{Effective QHAM Capacity using \textit{ibmq\_16\_melbourne}\cite{ibmq_16_melbourne} hardware. (a)} Majority vote accuracy for the QHAM in hardware for $n=\{3,4,5\}$, $m=\{2,3,4\}$, $\rho=0.2$ vs $\alpha = m/n$. \textbf{(b)} Density accuracy using the same test conditions as (a) vs $m/n$.}
\label{Fig9}
\end{figure}

\subsubsection*{QHAM Effective Memory Capacity in Hardware}
For limited cases, we are able to implement the QHAM in the \textit{ibmq\_16\_melbourne} hardware. The same test structure as used in simulation is performed for $n=\{3,4,5\}$, $m=\{2,3,4\}$, $\rho=0.2$ and 1024 shots, as shown in Fig. \ref{Fig9}. We note that the number of trials performed for these measurements is much lower than the number of trials performed in simulation due to hardware availability.

Clearly, the hardware implementation of the associative memory degrades the ability of the system to store memories, even more than the test with simulated noise from the noise model of the same hardware device. This is due to the significantly increased noise from lack of hardware connectivity and the increased number of qubits required for hardware execution of the QHAM circuit. While the noise model takes quantum error effects into account, it does not take into account these limitations. The noise model construction is explained in further depth in the Supplementary Information. Even for small memory cases such as the $n=\{3,4,5\}$ tests, the errors caused by the lack of full connectivity in the physical qubits become significant in the full system implementation. There is some recall in the single pattern case for $n=3$, but an accuracy metric of around 60 \% in the majority vote case is quite low for the production of meaningful results. In this case, quantum error correcting codes would become highly useful.

\section*{Discussion}

We have demonstrated a Quantum Hopfield Associative Memory system which can be executed in real quantum hardware via IBMQ. The number of qubits scales linearly with the length of the memory in simulation and in hardware where reset operations are available. If reset operations are not possible, the system also scales linearly in the number of updates. The accuracy of this system depends strongly on quantum errors and noise, but we demonstrate high resilience to these errors when extracting the quantum superposition states to the classical domain via a majority vote of many circuit shots. Noise modeling in simulation via IBMQ aids in understanding the effects of different types of noise on quantum circuit performance, while testing in hardware shows limitations due to a low number of qubits and limited connectivity. We benchmark the QHAM by measuring its effective memory capacity which is shown to mimic the effective memory capacity of classical Hopfield associative memory systems. The output of the QHAM can also remain in the quantum domain via $n$ entangled superposition states with high accuracy, which could then be used for further quantum computation.

The implementation of quantum associative memory models such as ours are significant for the advancement of machine learning in the domain of quantum computing. In the NISQ era of quantum computing, this successful QHAM implementation in hardware is a significant demonstration of what can be done in quantum hardware even in devices with a limited number of qubits and high noise. With applications in image recognition, optimization, data restoration from corrupt storage and more, our QHAM has high potential for further development in various application spaces. Further study pathways include noise-aware training, quantum algorithm developments for producing true Hopfield threshold dynamics in the quantum domain, and quantum hardware developments for increased connectivity. There have been promising developments in quantum associative memory models which theorize polynomial or exponential improvements in memory capacity,\cite{exponential_qham,qnn_quest,QuAM} which is a major motivation to continue this work to determine a route for true quantum advantage in this area. Our study demonstrates a step in this direction by being the first of its kind in which hardware implementation is explored.

\section*{Methods}

All QHAM testing was performed using IBMQ. Most data was collected via the QASM Simulator using noise models produced by the hardware systems, and some hardware tests were performed for cases which were small enough to fit on the quantum hardware, specifically on \textit{ibmq\_16\_melbourne}, the largest publicly available IBMQ hardware device at the time of this work with 15 qubits. The noise models are frequently updated and calibrated by IBM to maintain accuracy to the true systems, and new versions of hardware are frequently being released. In this study, the specific IBMQ hardware versions utilized were: \textit{ibmq\_5\_yorktown} (\textit{ibmqx2}) v2.2.6, \textit{ibmq\_16\_melbourne} v2.3.8, \textit{ibmq\_athens} v1.3.10, \textit{ibmq\_santiago} v1.3.10, \textit{ibmq\_lima} v1.0.2, \textit{ibmq\_quito} v1.0.4, \textit{ibmq\_belem} 1.0.0, and \textit{ibmq\_armonk} v2.4.0.

All QASM Simulator and hardware tests run with IBMQ use a variable called ``shots'' which represents the number of repeated tests which will be performed in the QASM simulator in order to gain an output probability distribution which is close to describing the true superposition state of the qubits. For the tests shown in Figs. \ref{Fig4}, \ref{Fig5} and \ref{Fig6} and Table S2, we use $shots=8192$, the maximum allowed by IBMQ at the time of these experiments. For the Monte Carlo tests shown in Figs. \ref{Fig7}, \ref{Fig8} and S2, we use $shots=1024$ to save on run time, then perform many of these same tests and take a separate majority vote of the results to produce the effective memory capacity measurements. For all $n$ in simulation we perform 1000 tests for the Monte-Carlo simulation (for $1024*1000=1,024,000$ total ``shots''), and in hardware we perform only 30 tests due to long hardware queue times.




\section*{Acknowledgements}

We acknowledge the use of IBM Quantum Services for this work. The views expressed are those of the authors, and do not reflect the official policy or position of IBM or the IBM Quantum team.

\section*{Author contributions statement}

NEM designed the quantum neuron and QHAM system, performed IBMQ simulations and hardware experiments, and analyzed the data presented in this work. SM advised the study and provided valuable insight into the machine learning concepts analyzed in this work. NEM wrote the manuscript and supplemental information text and prepared all figures. All authors reviewed the manuscript.

\section*{Additional information}

\textbf{Data and code availability:} The code used for IBMQ simulation, testing and data analysis and the data produced and analyzed in this study are available from the authors upon reasonable request.

\noindent\textbf{Competing interests:} The authors declare that there are no competing interests.


\end{document}


\flushbottom
\maketitle

\thispagestyle{empty}

\section*{Gate Complexity of the Quantum Neuron Designs}

One of the principal advantages of our proposed quantum neuron design as opposed to the quantum neuron design by Cao et al.\cite{GG_QN} is its improvement in the gate complexity of the neuron circuit. To quantify gate complexity, we consider the basis gates \{$CNOT$, $ID$, $R_z$, $SX$, $X$\}, which are the basis gates for IBMQ hardware.\cite{IBMQ} In the quantum neuron circuits in Figs. 1 and 2, we see the gates $CR_y$, $R_y$, $CY$, $R_z$, and $SWAP$ which must be decomposed to the basis gates to be executed in IBMQ hardware. The decomposition of these gates corresponds to $10$, $4$, $3$, $1$ and $3$ basis gates respectively.\cite{IBMQ_qasm} For example, the $CY$ gate decomposes to the basis gate sequence $R_z(-\frac{\pi}{2})\rightarrow CNOT\rightarrow R_z(\frac{\pi}{2})$ where the $R_z$ gates are performed on the target qubit of the $CY$ gate and the control and target of the $CNOT$ gate are the same as the original $CY$ gate. Note there are some instances where these numbers reduce, such as $CR_y(\pi)$ only requiring $8$ gates to perform due to simplifications permitted by the rotation angle $\pi$ as opposed to a more complex angle, but in general we will assume that these edge cases occur infrequently.

The Cao et al.\cite{GG_QN} neuron circuit shown in Fig. 1 requires $2(n-1)$ $CR_y$ rotations, two $R_y$ rotations for the $\beta$ terms, one $CY$, and one $R_z$ per update. If the update fails to produce the correct rotation on the output qubit, an additional $R_y$ rotation is used to reverse the rotation, and the primary construction is repeated $f$ additional times until a successful rotation is achieved. Once success is achieved, a $SWAP$ gate is used to overwrite the original qubit, and the ancilla qubit is reset if possible. This process occurs for each of the $u$ updates the system performs. Therefore, the gate complexity of this circuit in the IBMQ hardware basis gates is equal to $[20n(f+1)-4f-5]u$ or approximately $O(nuf)$, not including non-unitary operations such as measurement and reset. It is also worth noting the gate complexity in terms of $CNOT$ gates and single-qubit gates, as $CNOT$ gates are the more resource intensive of the standard basis gates. The RUS neuron requires $[16n(f+1)-f-5]u$ single qubit gates and $[4n(f+1)-3f]u$ $CNOT$ gates.

The gate complexity of the Cao et al.\cite{GG_QN} neuron also increases significantly if the circuit is recursively applied in order to create a rotation function with a sharper slope described by $R_y(2(arctan(tan^2\phi))^{\circ k})$ where $k$ describes the number of recursions. We benchmark our system against the $k=1$ system in this study, but it is also useful to compare the qubit and time overhead of the full RUS system. Increasing $k$ helps to decrease the error rate $\epsilon$ and increase the success rate $1-\nu$ of the RUS system. To simulate an associative memory of $n$ neurons with $t$ updates up to error $\epsilon$ and success probability at least $1-\nu$, with weights and biases discretized with resolution $\delta$, Cao et al. cite their expected runtime as $O(\frac{n^{2.075}t}{\delta^{2.075}\epsilon^{3.15}}log(\frac{t}{\nu}))$ and their qubit overhead as $O(n+log\frac{n}{\delta\epsilon^{1.75}})$.\cite{GG_QN}

\renewcommand{\thefigure}{S1}
\begin{figure*}[b]
\centering
\includegraphics[width=0.75\textwidth]{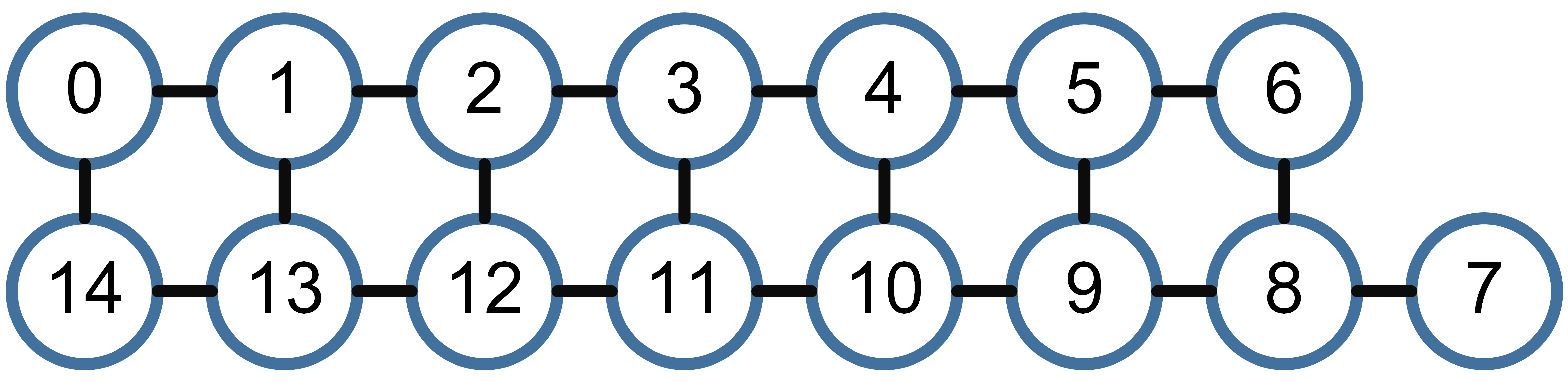}
\caption{\textbf{Melbourne hardware qubit architecture.} Hardware architecture of the IBMQ \textit{ibmq\_16\_melbourne} system, available from IBMQ.\cite{IBMQ,ibmq_16_melbourne}}
\label{FigS1}
\end{figure*}

For our proposed quantum neuron design, we cut the number of required $CR_y$ and $R_y$ gates in half as compared to the $k=1$ RUS neuron, remove the $CY$ and $R_z$ gates, and do not require repeating the circuit execution due to failure. Our gate complexity therefore reduces to $(10n-3)u$ total gate operations, or approximately $O(nu)$. This depth breaks down to $(8n-4)u$ single qubit gates and $(2n+1)u$ $CNOT$ gates, both significantly less than the gate complexity of the RUS neuron. Our neuron complexity is further improved by not requiring non-unitary reset and measurement operations or any classical conditioning, which the RUS neuron requires with each failure of the RUS construction. Additionally, Cao et al. propose performing many iterations of the same neuron in order to take a majority vote of the measurement results and reassign the neuron value according to the most frequently measured result. Since our design proceeds with all necessary updates without stopping to measure the output of each neuron mid-circuit execution, our design also requires significantly less repetition and measurement.

\renewcommand{\thetable}{S1}
\begin{table*}[t]
\caption{\textbf{Measured noise characteristics of IBMQ hardware.} Noise and error modeling characteristics of IBMQ hardware used to create their various noise models.\cite{ibmq_16_melbourne,ibmqx2,ibmq_lima,ibmq_quito,ibmq_belem,ibmq_athens,ibmq_santiago,ibmq_armonk}}
\label{TableS1}
\resizebox{\textwidth}{!}{
\begin{threeparttable}
\begin{center}
\begin{tabular}{ | c | B{1.45cm} | c | B{1.75cm} | B{1.75cm} | B{2.4cm} | B{2.9cm} | B{3.5cm} | B{1.4cm} | }
 \hline
 Device & Number of Qubits & Processor & Average $T_1$ Time ($\mu$s) & Average $T_2$ Time ($\mu$s) & Avg Readout Error Rate ($\%$) & Avg Single Qubit SX Error Rate ($\%$) & Avg Two Qubit CNOT Error Rate ($\%$) & Quantum Volume \\ 
 \hline
 ibmq\_16\_melbourne & 15 & Canary & 55.60 & 56.15 & 6.89 & 0.125 & 3.05 & 8 \\  
 \hline
 ibmqx2* & 5 & Canary & 59.30 & 36.05 & 4.58 & 0.099 & 1.75 & 8 \\ 
 \hline
 ibmq\_athens & 5 & Falcon & 74.08 & 91.22 & 2.02 & 0.045 & 1.21 & 32 \\ 
 \hline
 ibmq\_santiago & 5 & Falcon & 121.58 & 101.01 & 4.36 & 0.024 & 0.74 & 32 \\ 
 \hline
 ibmq\_lima & 5 & Falcon & 79.79 & 85.86 & 2.60 & 0.034 & 0.97 & 8 \\ 
 \hline
 ibmq\_quito & 5 & Falcon & 81.83 & 80.41 & 2.92 & 0.054 & 1.21 & 16 \\     
 \hline
 ibmq\_belem & 5 & Falcon & 75.62 & 100.24 & 2.56 & 0.026 & 1.19 & 16 \\ 
 \hline
 ibmq\_armonk & 1 & Canary & 138.19 & 222.74 & 2.60 & 0.019 & N/A & 1 \\
 \hline
\end{tabular}
\begin{tablenotes}
      \small
      \item *Also known as ibmq\_yorktown
\end{tablenotes}
\end{center}
\end{threeparttable}}
\end{table*}

\renewcommand{\thetable}{S2}
\begin{table*}[b]
\caption{\textbf{Noisy QHAM results using hardware and simulated noise models.} Comparison of results from hardware and simulation with various noise models \cite{ibmq_16_melbourne,ibmqx2,ibmq_lima,ibmq_quito,ibmq_belem,ibmq_athens,ibmq_santiago,ibmq_armonk} for the $n=4$ associative memory configuration tested in Fig. 6. In all cases, the only qubit which is chosen to be updated is Qubit 3.}
\label{TableS2}
\centering
\begin{threeparttable}
\begin{center}
\begin{tabular}{ | c | c | c | c | c | c | }
 \hline
 Device & Qubit 1 & Qubit 2 & Qubit 3 & Qubit 4 & Avg Accuracy ($\%$) \\ 
 \hline
 Target Result & 0.0 & 1.0 & 1.0 & 0.0 & 100.0 \\  
 \hline
 Noiseless Simulation & 0.0 & 1.0 & 1.0 & 0.0 & 100.0 \\ 
 \hline
 ibmq\_16\_melbourne Noise Model & 0.011 & 0.94 & 0.94 & 0.065 & 95.1 \\
 \hline
 ibmqx2* Noise Model & 0.024 & 0.95 & 0.93 & 0.023 & 95.8 \\
 \hline
 ibmq\_athens Noise Model & 0.0089 & 0.97 & 0.96 & 0.0052 & 97.9 \\
 \hline
 ibmq\_santiago Noise Model & 0.0095 & 0.98 & 0.97 & 0.068 & 96.8 \\ 
 \hline
 ibmq\_lima Noise Model & 0.0089 & 0.98 & 0.97 & 0.014 & 98.2 \\ 
 \hline
 ibmq\_quito Noise Model & 0.024 & 0.97 & 0.95 & 0.025 & 96.8 \\ 
 \hline
 ibmq\_belem Noise Model & 0.037 & 0.97 & 0.96 & 0.013 & 97.0 \\ 
 \hline
 ibmq\_16\_melbourne Hardware & 0.10 & 0.70 & 0.67 & 0.049 & 80.5 \\
 \hline
\end{tabular}
\begin{tablenotes}
      \small
      \item *Also known as ibmq\_yorktown
\end{tablenotes}
\end{center}
\end{threeparttable}
\end{table*}

\section*{IBMQ Melbourne Hardware Architecture}

The architecture of the \textit{ibmq\_16\_melbourne} \cite{IBMQ,ibmq_16_melbourne} system in IBMQ shown in Fig. \ref{FigS1} is a grid architecture, meaning that no single qubit is directly physically connected to any more than three other qubits. To route quantum gate operations between any qubits which are not directly connected, intermediary operations are used by the IBMQ Transpiler to route the operation through the connecting qubits from the control qubit to the target qubit. This routing often consists of many quantum SWAP gates, each composed of three CNOT gates, which can be used to swap the quantum states of any two connected qubits. For example, a CNOT gate implemented with qubit 0 as the control and qubit 2 as the target using the \textit{ibmq\_16\_melbourne} device shown in Fig. \ref{FigS1} would need to first route through qubit 1, causing the need for additional gates and thus a higher error rate than a comparable CNOT gate implemented on neighboring qubits.

\section*{IBMQ Noise Model Parameters}

We utilize noise models created by IBMQ to study the impact of quantum hardware noise on our neuron and QHAM. IBMQ lists average thermal relaxation ($T_1$) and dephasing ($T_2$) times\cite{Decoherence} for each qubit, average single-qubit gate error rates for each qubit (measured by the average error rate of the Sqrt-X ``SX'' gate), average two-qubit gate error rates for each two qubit control-target pattern (measured by the average error rate of the CNOT gate), and average readout errors of each qubit for all of its hardware devices.\cite{ibmq_16_melbourne,ibmqx2,ibmq_lima,ibmq_quito,ibmq_belem,ibmq_athens,ibmq_santiago,ibmq_armonk} Using these characteristics, noise models are created which are used to approximate errors resulting from operating circuits in hardware. A summary of parameters used in these noise models is shown in Table \ref{TableS1}. Overall, the combined error rates can be used to calculate a metric called quantum volume \cite{quantum_volume} which quantifies the largest random quantum circuit of equal width and depth that each system can successfully implement. Naturally, this metric tends to be larger for systems with low gate and readout error rates and more advanced backend processing. However, this metric does not take physical hardware connectivity into account, leading to significant differences in noise model simulation and true hardware results, as shown in Figs. 8 and 9. and Table \ref{TableS2} For much of our QHAM testing we choose to use \textit{ibmq\_16\_melbourne} because of its large number of physical qubits, but for smaller circuits it can be wise to use systems such as \textit{ibmq\_athens} or \textit{ibmq\_santiago} because of their larger quantum volume.

\section*{Noisy QHAM Simulation}
We perform the $n=4$, single update test of Figs. 6 and 4a using each of the available noise models to benchmark the performance of the update step with respect to noise, as shown in Table \ref{TableS2}. The average accuracy shown in the table is calculated as the average bitwise accuracy of the four qubits compared to their target value of $\ket{0110}$. It is clear that the final accuracy of the associative memory system depends greatly on the average error rates of the hardware. For example, \textit{ibmq\_16\_melbourne} and \textit{ibmqx2} are the systems with the lowest $T_1$ and $T_2$ times, highest readout error rates, and highest single-qubit and two-qubit gate error rates, and as expected, they show the worst overall accuracy for the results shown in Table \ref{TableS2}. However, it is interesting that \textit{ibmq\_lima} shows the highest average accuracy in this test, even though it has a lower quantum volume \cite{quantum_volume} than \textit{ibmq\_athens}, \textit{ibmq\_santiago}, \textit{ibmq\_quito} and \textit{ibmq\_belem}.

\renewcommand{\thefigure}{S2}
\begin{figure*}[!b]
\centering
\includegraphics[width=\textwidth]{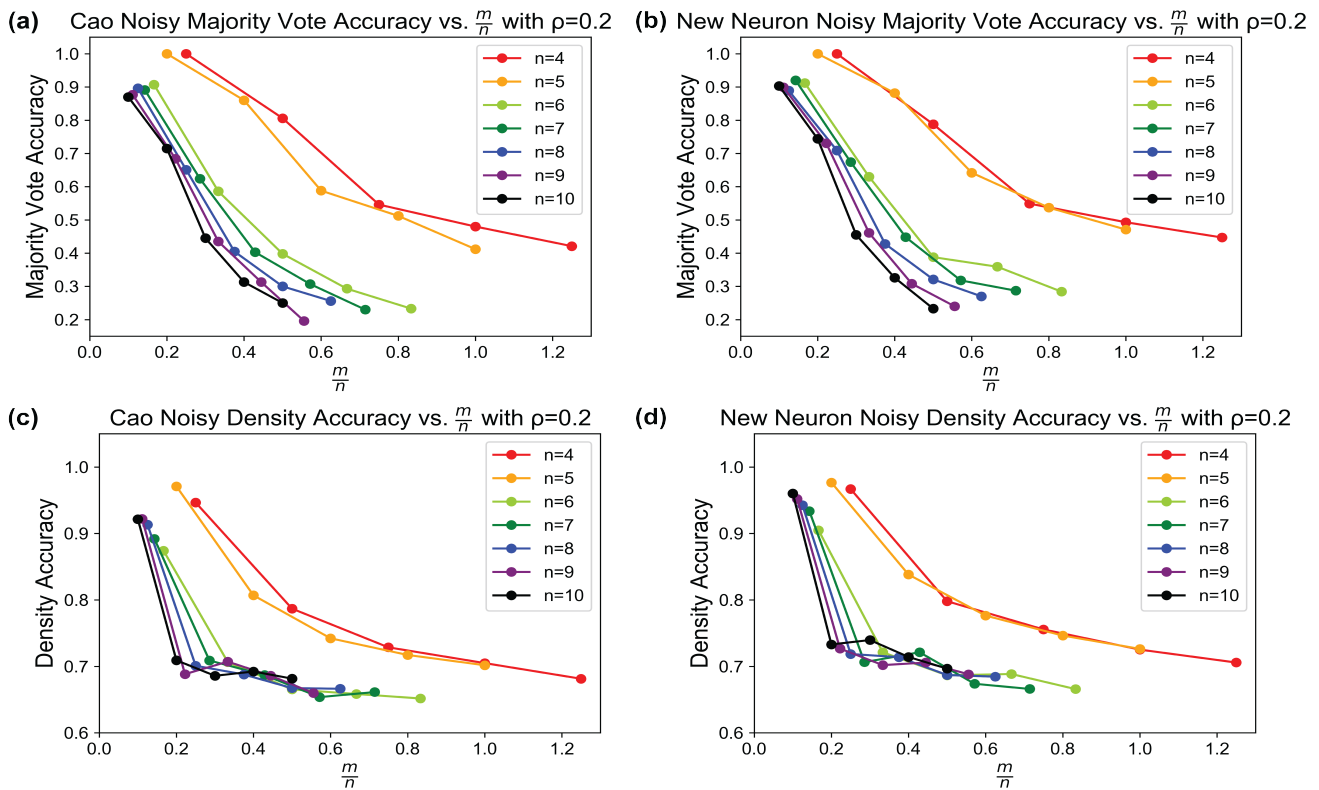}
\caption{\textbf{Effective memory capacity simulation of both neuron systems.} Simulations of the effective memory capacity of associative memories built from the Cao et al. \cite{GG_QN} neuron and our neuron using the \textit{ibmq\_quito} noise model following the same process as Figs. 7 and 8.}
\label{FigS2}
\end{figure*}

Since the qubit connectivity of the hardware system on which the QHAM is executed is crucial for performance, and as Fig. 6 shows, the accuracy of the QHAM in hardware can be worse than in simulation. For example, the \textit{ibmq\_16\_melbourne} hardware test shown in Fig. 6 clearly performs worse than simulations with the corresponding noise model. This is explained by the hardware execution of the circuit requiring additional operations and qubits than simulation. For the system we are modeling with a QHAM of length $n=4$ with a single update, we require 5 qubits which must all be connected to complete the update. However, there is no qubit in the \textit{ibmq\_16\_melbourne} hardware which is directly connected to four other qubits, so performing the desired update requires routing the control signals through multiple qubits to reach their targets. This process requires several SWAP gates, and errors from each operation tend to compound and increase the error rate. Thus, the hardware performance of our QHAM is highly dependent on the overall connectivity of the qubits.

\section*{Comparing the Effective Memory Capacity of Both Quantum Neuron Systems}
We benchmark our QHAM against an associative memory created using the quantum neuron by Cao et al. \cite{GG_QN} in order to understand how the addition of simulated noise (using the \textit{ibmq\_quito} noise model) affects both systems, as shown in Fig. \ref{FigS2}. The same process of testing effective memory capacity and tuning with the $u$ parameter as in Figs. 7 and 8 is implemented for $\rho=0.2$, $n\in [4,5,\cdots,10]$ and $m\in [1,2,3,4,5]$. Both networks have similar effective memory capacities which, in the majority vote accuracy metric, closely mirror the expected memory capacity from classical associative memories as described by equation 9. As we expected from the results of Fig. 3, our network shows a minor improvement in both majority vote accuracy and density accuracy in nearly all cases of around 5 \%. Even though our neuron response in Fig. 3 is flatter than the Cao et al. neuron response, our network still sees this minor improvement due to its resiliency to noise caused by a lower gate complexity.
